\documentclass[a4paper,12pt]{article}
\usepackage{geometry}
\usepackage[numbers,sort&compress]{natbib}
\usepackage{graphicx}
\usepackage{amssymb}
\begin{document}
\title{Probability distribution function of dipolar field in two-dimensional spin ensemble}
\author{Andrey V. Panov}
\date{}
\maketitle
\begin{center}
\noindent{\textit{Institute of Automation and Control Processes,\protect\\
Far East Branch of Russian Academy of Sciences,\protect\\
5, Radio st., Vladivostok, 690041, Russia}\\
\texttt{panov@iacp.dvo.ru}}
\end{center}

\begin{abstract}
We theoretically determine the probability distribution function of the net field of the random planar structure of dipoles which represent polarized particles. At small surface concentrations $c$ of the point dipoles this distribution is expressed  in terms of special functions. At the surface concentrations of the dipoles as high as $0.6$ the dipolar field obey the Gaussian law. To obtain the distribution function within transitional region $c<0.6$, we propose the method based on the cumulant expansion. We calculate the parameters of the distributions for some specific configurations of the dipoles. The distribution functions of the ordered ensembles of the dipoles at the low and moderate surface concentrations have asymmetric shape with respect to distribution medians. The distribution functions allow to calculate various physical parameters of two-dimensional interacting nanoparticle ensembles.

The final publication is available at www.epj.org, DOI: 10.1140/epjb/e2012-30057-7
\end{abstract}

A considerable interest in studying the electric, magnetic and optical
properties of nanocomposites is connected today with two-dimensional systems containing particles. These nanoparticles may be polarized electrically or magnetically and thus to a first approximation they can be regarded as dipoles. Since sufficiently high particle concentrations are widely encountered in practice, the long-range dipole-dipole interparticle interaction should be taken into account. In particular, the dipolar ferromagnetism was experimentally observed in monolayer nanoparticle arrays \cite{Yamamoto08,Yamamoto11}. Moreover, the dipole-dipole interactions are substantial for the optical properties of nanocomposites \cite{Sersic09,Liu11,Panov10}.

The consideration of the dipole-dipole interactions is a complex many-body problem due to their long-range and anisotropic behavior. Frequently, this problem is solved using first principle numerical simulations. But such a direct computation of the dipole-dipole interactions in an ensemble is a resource consuming problem, so we can use the probability distribution function of dipolar interaction field in order to allow for interparticle interactions  and to derive some analytical expressions. After obtaining the probability distribution function, it is possible to calculate some physical parameters of a system by means of the methods of the statistical mechanics.

As was shown before, in three-dimensional sample at the low volume concentrations of dipoles (spins), the distribution function of the net dipolar field is Cauchy-Lorentzian \cite{Holtsmark19,Grant64,Klein68}, at the large concentrations the distribution becomes Gaussian \cite{Grant64}. The condition for transition to the normal law was shown in Ref.~\cite{Shcherbakov79:eng}. Because of the different geometry, these results cannot be applied to the two-dimensional systems. In this study,
 we determine the probability distribution function of the random dipolar field of the flat structure of the spins.

Let us consider the system of circular particles (nanoislands) which lie randomly in a plane. These particles are assumed to have the dipole polarization. We suppose that the particles are surrounded by dielectric matrix. At the coordinate origin, net dipolar field projection $E$ onto a selected direction falls within range $(E,E+dE)$ with probability 
\[
\delta \left[E - \sum^N_{k = 1}\xi_{k} \right] d E ,
\]
where $\xi_{k} \left( \mathbf{p}_k,\mathbf{r}_k \right)$ is the
field induced by $k$-th dipole $\mathbf{p}_k$ located at position $\mathbf{r}_k$, $\delta$ is Dirac's delta function. Here, $E$ represents the projection of either electric or magnetic field. Then the random field distribution function will be
\begin{equation}
W (E ) d E = \int \delta \left[E - \sum^N_{k =
1}\xi_k \right]\prod^N_{k =
1} \tau_k\left( \mathbf{p}_k \right)\, d \mathbf{p}_{k}\, d\mathbf{r}\, d E ,
\end{equation}
where $\tau_k \left( \mathbf{p}_k \right)$ are the distribution functions for dipole moments, $d\mathbf{r}=r\,dr\,d\varphi$. After applying Markov's method \cite{Chandra43}, under the assumption that all the $\tau_k$ distributions are identical $\tau_k \left( \mathbf{p}_k \right)=\tau\left( \mathbf{p}\right)$, $\mathbf{p}_k=\mathbf{p}$,  we obtain characteristic function $A(\rho)=\exp\left[-C(\rho) \right] $  with
\begin{equation}
 C(\rho)=\frac{c}{s}\int\left[ 1-\exp\left(i\rho \xi\right) \right] \tau\left( \mathbf{p}\right)
 d \mathbf{p}\, d\mathbf{r},
 \label{initint}
\end{equation}
where  $s$ is the area of the particle, $c=N s/S$ is the surface concentration of the dipolar polarized particles, $S$ is the area of the sample. It is worth mentioning that $C(\rho)$ is the cumulant-generating function with opposite sign.
For the dipolar interaction
\begin{equation}
 \xi=\frac{3(\mathbf{r}\cdot\mathbf{p})\mathbf{r}-\mathbf{p}r^2}{\varepsilon_m r^5}=\frac{D}{r^3},
\end{equation}
where $\varepsilon_m$ is the dielectric function of surrounding matrix in the case of interacting electric dipoles. 

First, we give a consideration to the case of point dipoles which corresponds to limit $c\rightarrow 0$. Upon substituting $y=1/r^3$, $y_1=1/R^3$, $R$ is the radius of the sample, and neglecting the size of the particles,  Eq.~\ref{initint} becomes
\begin{equation}
 \int\int_{y_1}^\infty{\frac {\tau c\left( 1-\exp(i\rho D y)
 \right) dy }{3y^{5/3}s}}d \mathbf{p}\,d\varphi=
 \int\frac{\tau c}{2s}\left\lbrace \frac{1 - \exp(i y_1 \rho D)}{y_1^{2/3}} -
 \frac{i\rho D \Gamma  \left( 1/3,-i y_1 \rho D \right) }{(-i\rho D)^{1/3}}
 \right\rbrace \,d\mathbf{p}\,d\varphi,
 \label{charfull}
\end{equation}
where $\Gamma$ denotes the incomplete gamma function.
For the small values of $y_1$ Eq.~\ref{charfull} takes form
\begin{equation}
 C(\rho)=\left| \rho \right|^{2/3}{3}^{2/3}(B_1 + i B_2\mathop{\mathrm{sign}}\rho) - i\rho E_0
 \label{surfchar}
\end{equation}
with scale parameter
\begin{equation}
 B_1=\frac{1}{2}\frac {\pi c}{3^{7/6}\Gamma\left( 2/3 \right)s}\int\tau\left| D \right|^{2/3} \,d\mathbf{p}\,d\varphi,
\end{equation}
skewness
\begin{equation}
 B_2=-\frac{1}{2}\frac {\pi c}{3^{2/3}\Gamma\left( 2/3 \right)s}\int\tau\left| D \right|^{2/3}\mathop{\mathrm{sign}}D \,d\mathbf{p}\,d\varphi
\end{equation}
and median
\begin{equation}
 E_0=-\frac{c}{s}\int \tau  D y_1^{1/3} \,d\mathbf{p}\,d\varphi.
\end{equation}
It should be noticed that $B_2$ and $E_0$ may become zero due to the alternating nature of the dipolar fields, for example, when one half of the spins is compensated by another part with the opposite direction of the dipoles. On this occasion the system has no ordering.

The obtained characteristic function corresponds to a stable distribution. It was shown in Ref.~\cite{Uchaikin99} that for the infinite two-dimensional ensemble of point sources with some types of field dependencies $\propto 1/r^\mu$ the characteristic function is proportional to $\exp(-|\rho|^{2/\mu})$. Our result correlate with their findings. For  the special case of Ising dipoles lying in $x$-$y$ plane, the similar characteristic function was derived in Ref.~\cite{Meilikhov04}.%
\footnote{An attempt to get the characteristic function in [L.L.~Afremov. Dissertation of doctor of sciences, Vladivostok, Far-Eastern state university (1999)] was unsuccessful due to errors.}
In order to apply inverse Fourier transform, we split integration over $\rho$ into two ranges for negative and positive $\rho$. Designating $B=B_1+i B_2$, $\tilde{E}=E-E_0$,  $x=B/|\tilde{E}|^{2/3}$ we obtain distribution function:
\begin{eqnarray}\nonumber
 W(E)&=&\left\lbrace \frac{i}{3}\mathop{\mathrm{sign}}\tilde{E}\left[ x\mathop{\mathrm{Bi}}x^2+\mathop{\mathrm{Bi}}\nolimits'x^2\right] -x\mathop{\mathrm{Ai}}x^2 -\mathop{\mathrm{Ai}}\nolimits'x^2\right\rbrace \frac{x}{|\tilde{E}|}\exp\frac{2x}{3} - {}
 \\ \nonumber & &
 \left\lbrace \frac{i}{3}\mathop{\mathrm{sign}}\tilde{E}\left[ x^*\mathop{\mathrm{Bi}}x^{*2}+\mathop{\mathrm{Bi}}\nolimits'x^{*2}\right] +x^*\mathop{\mathrm{Ai}}x^{*2} +\mathop{\mathrm{Ai}}\nolimits'x^{*2}\right\rbrace \frac{x^*}{|\tilde{E}|}\exp\frac{2x^*}{3}+{} \\ & &
 \frac{i}{2\pi\tilde{E} }\left\lbrace {}_2 F_2\left(\frac{1}{2},1;\frac{1}{3},\frac{2}{3};x^*\right)- {}_2 F_2\left(\frac{1}{2},1;\frac{1}{3},\frac{2}{3};x\right)\right\rbrace ,
 \label{surfdipdist}
\end{eqnarray} 
where $\mathop{\mathrm{Ai}}$, $\mathop{\mathrm{Bi}}$, $\mathop{\mathrm{Ai}}\nolimits'$,  $\mathop{\mathrm{Bi}}\nolimits'$ are the Airy functions and their derivatives, ${}_2 F_2$ is the generalized hypergeometric function, the asterisk denotes complex conjugation. In the case of the small values of $\tilde{E}$ where the computation of the special functions involves difficulties, the characteristic function can be expanded into a series and the integral over $\rho$ can be taken separately. It gives formula:
\begin{equation}
W(E)=\frac{1}{384\sqrt{\pi}}\mathop{\mathrm{Re}}\left\lbrace \frac{48B^3-35\tilde{E}^2}{B^{9/2}}-\frac{64 i \tilde{E}}{\sqrt{\pi}B^3}+
\frac{48B^{*\,3}-35\tilde{E}^2}{B^{*\,9/2}}-\frac{64 i \tilde{E}}{\sqrt{\pi}B^{*\,3}}\right\rbrace.
\end{equation} 
Distribution (\ref{surfdipdist}) is asymmetric with respect to its median. For the spin arrangement without ordering $B_2=0$ and $E_0=0$, so function~(\ref{surfdipdist}) becomes symmetric and simplifies to
\begin{equation}
 W(E)=-\frac{2B}{|{E}|^{5/3}}\left\lbrace \mathrm{Ai}'\frac{B^2}{|{E}|^{4/3}}+\frac{B}{|{E}|^{2/3}}\mathrm{Ai}\frac{B^2}{|{E}|^{4/3}}\right\rbrace \exp\frac{2B^3}{3{E}^2}.
 \label{surfdipsymdist}
\end{equation}
In this specific case, the distribution was expressed through the Whittaker function in  Ref.~\cite{Uchaikin99} before:
\begin{equation}
 W(E)=\frac{1}{2\sqrt{3\pi}|{E}|}\exp\left(\frac{2B^3}{3{E}^2} \right)
 W_{-\frac12,\frac16} \left(\frac{4B^3}{3{E}^2}\right).
 \label{surfdipwhittaker}
\end{equation}
Both Eq.~\ref{surfdipsymdist} and Eq.~\ref{surfdipwhittaker} are equivalent but the first form is more computationally efficient.
In the vicinity of zero (${E}\rightarrow0$), Eq.~\ref{surfdipsymdist} is approximated by formula
\begin{equation}
 W(E)=\frac{48 B^2-35{E}^2}{192\sqrt{\pi}B^{9/2}}.
\end{equation} 

It should be emphasized that Eq.~\ref{surfdipdist} with alternative parameters $B$ and $E_0$ also describes the distribution of interaction fields $\propto 1/r^{4}$ in the three-dimensional space.

%To take into account the finiteness of the particle in the first approximation, we proceed as follows: let us replace in the right hand side of Eq.~\ref{charfull} $y_1$ with $y_0=1/(2r_0)^3$, $r_0$ is the radius of the particle, $2r_0$ is the minimal distance between particle centers. The leading non-oscillating term at $y_0\rightarrow \infty$ of resulting expression is
%$
%\frac{(2r_0)^2c}{2s}
%$ (integration of $\int \tau\,d\mathbf{p}\,d\varphi$ yields unity). This term must be subtracted from formula (\ref{surfchar}). Further it gives factor $\exp\left( {2r_0^2c}/{s}\right) $ to Eq.~\ref{surfdipdist}. The similar multiplier for Cauchy distribution was obtained in Ref.~\cite{Grant64}.

\begin{figure}
 {\centering
 \includegraphics[width=7.5cm,keepaspectratio=true]{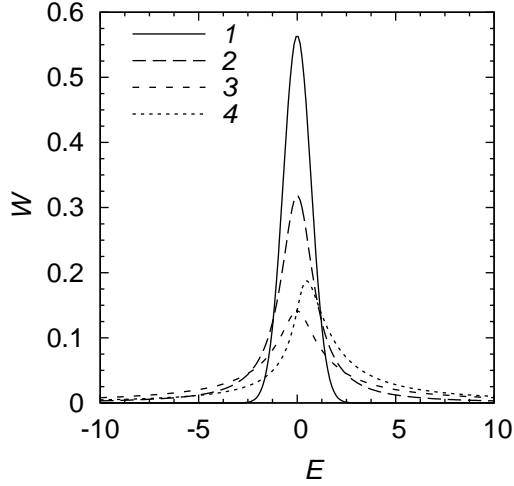}\par}
 
\caption{Gaussian (1), Cauchy (2),  Eq.~\ref{surfdipsymdist} (3) and Eq.~\ref{surfdipdist} (4) distributions with a unit area under the curves and zero medians. Curve 4 is calculated using $B=0.817-0.333i$.\label{fig:surfdist}}
\end{figure} 

Fig.~\ref{fig:surfdist} shows functions~(\ref{surfdipdist}) and (\ref{surfdipsymdist}) in comparison with the Cauchy and Gaussian distributions. As well as the Cauchy law, distribution~(\ref{surfdipdist}) has no moments. As one can see from Fig.~\ref{fig:surfdist} distribution~(\ref{surfdipsymdist}) is more flat than the Cauchy law. This fact is attributed to the faster reduction of $\sum \xi$ with $r$ in two dimensions and the increasing contribution of neighbor spins. The main feature of function~(\ref{surfdipdist}) is its asymmetry in contrast to the symmetric three-dimensional distribution of the point dipole fields. In fact, the latter arises from the fortunate circumstances when $C(\rho)$ depends only on the integer powers of $\rho$.

Below, we provide the distribution parameters for some spin configurations with identical $|\mathbf{p}|=p$. When $N\alpha$ dipoles are oriented perpendicularly to the plane and $N\beta$ dipoles have opposite direction (transverse Ising model):
\begin{equation}
\tau(\gamma)={\alpha\delta(\gamma)+\beta\delta(\gamma-\pi)},
\end{equation}
\begin{equation}
 B_\perp=\frac{\pi^2 c p^{2/3}}{3^{7/6}\varepsilon_m^{2/3} s\Gamma\left( 2/3\right) }\left( 1+i\sqrt{3}m\right) \approx \frac{c p^{2/3}(2.023+3.504i m)}{\varepsilon_m^{2/3}s},
\end{equation}
\begin{equation}
 E_{0\perp}=\frac{2\pi c p m}{R\varepsilon_m s},
\end{equation} 
where $m=\alpha-\beta$ is the order parameter, $\gamma$ is the polar angle of $\mathbf{p}$.

For dipoles lying in the plane and having two possible opposite directions:
\begin{equation}
 B_\parallel=\frac{\pi c p^{2/3}}{3^{7/6}\varepsilon_m^{2/3} 2 s\Gamma\left( 2/3\right) }(I_1-I_2\sqrt{3}i m)\approx \frac{c p^{2/3}(1.938-1.369i m)}{\varepsilon_m^{2/3} s},
\end{equation}
\[
I_1=\int\limits_0^{2\pi}\left|3\cos^2\varphi-1\right|^{2/3}\,d\varphi\approx6.019,\]
\[
I_2=\int\limits_0^{2\pi}\left|3\cos^2\varphi-1\right|^{2/3}\mathop{\mathrm{sign}}(3\cos^2\varphi-1)\,d\varphi\approx 2.455,
\]
\begin{equation}
 E_{0\parallel}=-\frac{\pi c p m}{R\varepsilon_m s}.
\end{equation} 

When the spins with zero transverse component are arbitrarily oriented in the plane:
\begin{equation}
\tau(\gamma)=\frac 1{2\pi},
\end{equation}
\begin{equation}
 B=\frac{c p^{2/3}}{2^{5/3}3^{7/6}\varepsilon_m^{2/3} s\Gamma\left( 2/3\right) }\int\limits_0^{2\pi}\int\limits_0^{2\pi}\left|3\cos(2\varphi-\gamma)+\cos\gamma\right|^{2/3}\,d\varphi\,d\gamma\approx 3.828\frac{c p^{2/3}}{\varepsilon_m^{2/3} s},
\end{equation}
\[
 E_{0}=0.
\]
These parameters are calculated under the assumption of homogeneous surrounding matrix. For other configurations of ambient media they may have more complicated dependence on $\varepsilon_m$.

%In action averaging using Eq.~\ref{surfdipdist} will cause divergences at $|E|\rightarrow\infty$. In that case the value of $E$ may be limited by finite values from possible nearest neighbor dipoles as proposed in Ref.~\cite{Chandra43} for the Holtsmark law.

Further, we investigate the transition to the normal law. At large dipole concentrations $c$, according to the central limit theorem, the stable distribution for the dipolar field switches to Gaussian. This can be shown expanding Eq.~\ref{initint} into series in terms of $\rho$ up to $\rho^2$. This expansion is valid for high $c$. The projection of net dipolar field onto a selected direction has probability distribution function
\begin{equation}
 W_G(E) = \frac{\exp(-(E+E_0)^2/2\sigma^2)}{\sqrt{2\pi}\sigma}
\label{gaussian}
\end{equation}
with
\begin{equation}
 E_0=-\frac{c}{s}\int \xi \tau(\mathbf{p})d \mathbf{p}\, d\mathbf{r}
 \label{defE0gauss}
\end{equation} 
and variance
\begin{equation}
 \sigma^2=\frac{c}{s}\int \xi^2 \tau(\mathbf{p})d \mathbf{p}\, d\mathbf{r}.
 \label{defsigmagauss}
\end{equation} 
The integration in Eqs.~\ref{defE0gauss} and \ref{defsigmagauss} over $r$ begins from minimal distance between the particles $2r_0$, $r_0$ is the radius of the particle.
These parameters are readily obtainable for the specific configurations of the finite particles. For the Ising model with the transverse alignment of the spins:
\begin{equation}
 E_{0\perp}=\frac{\pi c p m}{r_0\varepsilon_m s},\quad
 \sigma_\perp^2=\frac{\pi c p^2}{2s(2r_0)^4\varepsilon_m^2}.
\end{equation} 
For two possible opposite orientations of the dipoles in the plane:
\begin{equation}
 E_{0\parallel}=-\frac{\pi c p m}{2r_0\varepsilon_m s},\quad
 \sigma_\parallel^2=\frac{11\pi c p^2}{16s(2r_0)^4\varepsilon_m^2},
\end{equation} 
and on occasion of the randomly aligned dipoles lying in the plane:
\begin{equation}
 E_{0}=0,\quad
 \sigma^2=\frac{5\pi c p^2}{8s(2r_0)^4\varepsilon_m^2}.
\end{equation} 
It is particularly important that the negative sign of $E_{0\parallel}$ permits the in-plane dipolar ferromagnetic ordering \cite{surfmagn}.

Making use of the proposed in Ref.~\cite{Shcherbakov79:eng} comparison of $\sigma^2$ and the higher moments of spin arrangement function $\tau(\mathbf{p})$, we can evaluate the concentration at which $W$ becomes Gaussian:
\[
\frac{{c}\int \xi^3 \tau(\mathbf{p})d \mathbf{p}\, d\mathbf{r}}{3s\sqrt{N}\sigma^3} \ll 1, \quad \frac{{c}\int \xi^4 \tau(\mathbf{p})d \mathbf{p}\, d\mathbf{r}}{12 s N\sigma^4} \ll 1.
\] 
Such an estimate gives that this transition occurs at $c>0.5$ for the transverse alignment of the spins and at $c>0.6$ for the in-plane arrangement of the dipoles. These values are much greater than $0.1{-}0.15$ obtained for the three-dimensional sample \cite{Shcherbakov79:eng}. Again, it can be accounted for the fast saturation of the dipole sums in two dimensions and the relatively large contribution of neighbors which differs from the normal law.

\begin{figure*}
 \centering
 \includegraphics[width=7.5cm,keepaspectratio=true]{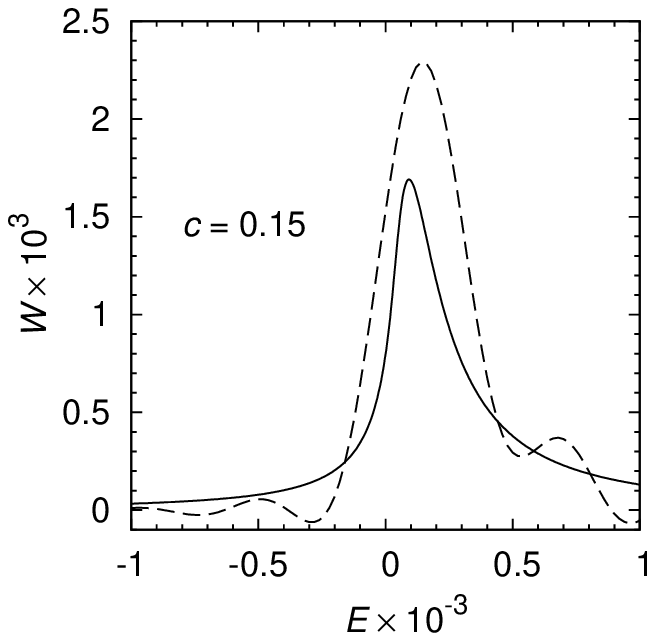}
 \includegraphics[width=7.5cm,keepaspectratio=true]{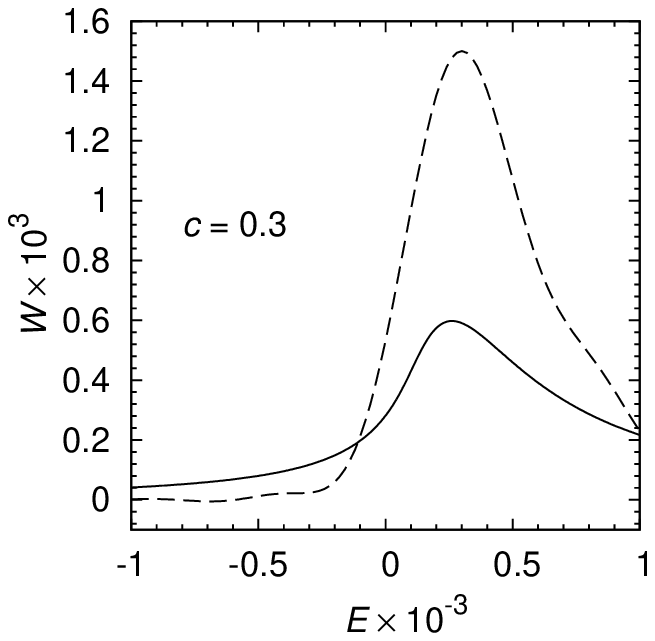}
\caption{Distributions of the net dipolar field of point dipoles (solid lines), Eq.~\ref{surfdipdist}, and finite dipoles calculated using expansion (\ref{Crhoexpan}) in characteristic function up to the 14th degree (dashed lines). The geometrical and physical parameters observed in Ref.~\citep{Yamamoto11} are used for the simulations.\label{fig:pdftransistional}}
\end{figure*}

From the above, we can see that stable distributions (\ref{surfdipdist}) and (\ref{gaussian}) are reached for the two extreme cases of the concentrations.
Within range of concentrations $0<c \lessapprox 0.6$, it is possible to add more terms of expansion  (\ref{initint}) in powers of $\rho$ (the cumulant expansion with opposite sign)
\begin{equation}
 C(\rho)=E_0\rho+\frac{1}{2}\sigma^2\rho^2+\frac{1}{3!}\mu_3\rho^3-\frac{1}{4!}\mu_4\rho^4+\frac{1}{5!}\mu_5\rho^5+\ldots,
\label{Crhoexpan}
\end{equation} 
where the moments are defined by $\mu_n=\frac{c}{s}\int \xi^n \tau(\mathbf{p})d \mathbf{p}\, d\mathbf{r}$ and make inverse Fourier transform numerically. As well as for $E_0$ and $\sigma^2$, it is easy to obtain expressions for the higher moments of the simple spin configurations. It is essential that the odd-numbered moments are proportional to order parameter $m$ and even-numbered $\mu_n$ do not depend on $m$ due to identity $\alpha+\beta=1$. The calculations of the distribution functions within the transitional range of the concentrations ($c=0.15$ and $c=0.3$) for in-plane Ising model with $m=1$ are depicted in Fig.~\ref{fig:pdftransistional}. As well as function (\ref{surfdipdist}), the negative cumulant expansion gives the asymmetric distribution in the ordered state when the odd moments are non-zero. The use of expansion (\ref{Crhoexpan}) up to the 14th degree of $\rho$ shows oscillations of the distribution function which are more pronounced for the lower concentrations. For comparison, Fig.~\ref{fig:pdftransistional} illustrates the distributions of the net point dipole fields calculated with the same parameters. We must remember that the latter curves are good approximation only for $c\rightarrow 0$. The employment of expansion (\ref{Crhoexpan}) benefits when moments $\mu_n$ depend on a small parameter which can be applied for further expansion. It is obvious from Fig.~\ref{fig:pdftransistional} that distribution~(\ref{surfdipdist}) differ significantly from the results of the negative cumulant expansion at higher $c$. From this it can be concluded that the model of the point sources is not valid at the moderate concentrations.

It should be underlined that the method based on the negative cumulant expansion is not restricted to the two-dimensional geometry and, even, to the dipolar field. This expansion may be useful for studying other types of interactions in ensembles with the easy calculation of $\mu_n$.

In conclusion, we expressed the distribution function of the random dipolar field in the two-dimensional structure in the limit of the low concentrations of dipoles  in terms of special functions. This distribution is gently sloping as compared with the Cauchy law obtained for three-dimensional sample. At rather high concentrations of polarized nanoparticles, the net random dipolar field adhere to the Gaussian law. Within the transitional range of the concentrations, the probability distribution function may be calculated after the expansion of the negative logarithm of the characteristic function in terms of the spin arrangement function moments. The results of this work may be utilized for the calculation of electromagnetic parameters of the planar structures with the dipole-dipole interactions.

\bibliography{inter_part}

\begin{thebibliography}{13}

\bibitem{Yamamoto08}
K.~Yamamoto, S.A. Majetich, M.R. McCartney, M.~Sachan, S.~Yamamuro,
  T.~Hirayama, Appl. Phys. Lett. \textbf{93}(8), 082502 (2008)

\bibitem{Yamamoto11}
K.~Yamamoto, C.R. Hogg, S.~Yamamuro, T.~Hirayama, S.A. Majetich, Appl. Phys.
  Lett. \textbf{98}(7), 072509 (2011)

\bibitem{Sersic09}
I.~Sersic, M.~Frimmer, E.~Verhagen, A.F. Koenderink, Phys. Rev. Lett.
  \textbf{103}, 213902 (2009)

\bibitem{Liu11}
J.Q. Liu, M.D. He, S.~Chen, C.P. Huang, L.~Zhou, Y.Y. Zhu, J. Phys.: Condens.
  Matter. \textbf{23}, 215303 (2011)

\bibitem{Panov10}
A.V. Panov, Opt. Lett. \textbf{35}(11), 1831 (2010)

\bibitem{Holtsmark19}
J.~Holtsmark, Annalen der Physik \textbf{363}(7), 577 (1919)

\bibitem{Grant64}
W.J.C. Grant, M.W.P. Strandberg, Phys. Rev. \textbf{135}(3A), A715 (1964)

\bibitem{Klein68}
M.W. Klein, Phys. Rev. \textbf{173}(2), 552 (1968)

\bibitem{Shcherbakov79:eng}
V.P. Shcherbakov, Fiz. Met. Metalloved. \textbf{48}(6), 1134 (1979)

\bibitem{Chandra43}
S.~Chandrasekhar, Rev. Mod. Phys. \textbf{15}(1), 1 (1943)

\bibitem{Uchaikin99}
V.V. Uchaikin, V.M. Zolotarev, \emph{Chance and Stability: Stable Distributions
  and their applications} ({VSP}, 1999), ISBN 9067643017

\bibitem{Meilikhov04}
E.Z. Meilikhov, R.M. Farzetdinova, J. Magn. Magn. Mater. \textbf{268}(1-2), 237
  (2004)

\bibitem{surfmagn}
A.V. Panov, Appl. Phys. Lett. \textbf{100}(5), 052406 (~3) (2012)

\end{thebibliography}

\end{document}